\documentclass[pra,reprint,nofootinbib,superscriptaddress]{revtex4-1}

\usepackage{graphicx}
\usepackage{dcolumn}
\usepackage{bm}
\usepackage{amsmath}
\usepackage{amssymb}
\usepackage{braket}
\usepackage{booktabs}
\usepackage{multirow}
\usepackage{setspace}
\usepackage{afterpage}
\usepackage{floatrow}
\usepackage{enumitem}
\usepackage{xr}

\begin{document}

\title{Magnetic hysteresis of a superconducting microstrip resonator with a high edge barrier}

\author{Sangil Kwon}
\email{kwon2866@gmail.com}
\affiliation{Institute for Quantum Computing, University of Waterloo, Waterloo, Ontario N2L 3G1, Canada}
\affiliation{Department of Physics and Astronomy, University of Waterloo, Waterloo, Ontario N2L 3G1, Canada}

\author{Yong-Chao Tang}
\affiliation{Institute for Quantum Computing, University of Waterloo, Waterloo, Ontario N2L 3G1, Canada}
\affiliation{Department of Electrical and Computer Engineering, University of Waterloo, Waterloo, Ontario N2L 3G1, Canada}

\author{Hamid~R.~Mohebbi}
\affiliation{High Q Technologies LP, Waterloo, Ontario N2L 0A7, Canada}

\author{David G. Cory}
\affiliation{Institute for Quantum Computing, University of Waterloo, Waterloo, Ontario N2L 3G1, Canada}
\affiliation{Department of Chemistry, University of Waterloo, Waterloo, Ontario N2L 3G1, Canada}
\affiliation{Perimeter Institute for Theoretical Physics, Waterloo, Ontario N2L 2Y5, Canada}
\affiliation{Canada Institute for Advanced Research, Toronto, Ontario M5G 1Z8, Canada}

\author{Guo-Xing Miao}
\affiliation{Institute for Quantum Computing, University of Waterloo, Waterloo, Ontario N2L 3G1, Canada}
\affiliation{Department of Electrical and Computer Engineering, University of Waterloo, Waterloo, Ontario N2L 3G1, Canada}

\date{\today}

\begin{abstract}
We investigate the magnetic hysteresis of a superconducting microstrip resonator with a high edge barrier.
We measure the magnetic hysteresis while either sweeping a magnetic field or tuning the edge barrier by high microwave current.
We show that the magnetic hysteresis of such a device is qualitatively different from that of one without an edge barrier and can be understood based on the generalized critical-state model.
In particular, we propose and demonstrate a simple and intuitive method that relies on a plot of the quality factor versus the resonance frequency for revealing the physical processes behind those hysteretic behaviors.
Based on this, we find that the interplay between the Meisser current and vortex pinning is essential for understanding the magnetic hysteresis of such a device.
\end{abstract}

\maketitle

\section{Introduction}
\label{sec:intro}

Metastability is one of the central pieces characterizing macroscopic properties of superconductors \cite{leggett}.
From a practical point of view, vortex pinning and a surface/edge barrier are particularly important sources of metastability.
They can be helpful for reducing magnetic field dependent losses by delaying vortex penetration (surface/edge barrier) and suppressing current-induced vortex motion (vortex pinning) \cite{matsushita}.
Their importance is more evident for a planar superconducting device because thin films are effectively strong type-II superconductors regardless of their Ginzburg--Landau parameter \cite{tinkham}.

The most well-known consequence of metastability in type-II superconductors is magnetic hysteresis in $M$-$H$ curves.
The hysteresis of a bulk material is mainly governed by vortex pinning \cite{matsushita}, while the hysteresis due to a surface/edge barrier becomes pronounced in a mesoscopic superconductor \cite{deo1999, geim2000, peeters2002, oxford}.
Following this tradition, the magnetic hysteresis of a planar superconducting device has been attributed to vortex pinning based on Bean-type critical state models \cite{meservey1969, lancaster, frunzio2005, deGraaf2012, bothner2012b, bothner2017}.

This approach, however, neglects the fact that most of the applied current in a planar superconducting device flows along the edges where the Meissner current is maximized.
Because of this high current density, the effects of the Meissner current on the magnetic properties of such a device are amplified and can be comparable to those of vortex pinning, especially if there is a reasonably high edge barrier \cite{kwon1}.
Although a series of theoretical studies based on the generalized critical-state model predict qualitatively different behaviors from those without an edge barrier \cite{maksimov1995, maksimov1997, elistratov1997, maksimov1998, elistratov2000, maksimova2001}, there has been little experimental study on this with a superconducting device.

In this work, we explore magnetic hysteresis in a superconducting microstrip resonator with a high edge barrier.
We show that the predictions from the generalized critical-state model are consistent with our experimental observations.
In addition, we propose a plot of $Q$ vs. $f^{-2}$, where $Q$ and $f$ are the quality factor and the resonance frequency, as a characterization method for the magnetic hysteresis of a superconducting microstrip resonator.
The motivation of this plot is to represent the characteristic relation between the real and imaginary parts of the complex resistivity for each contribution---either quasiparticle generation or vortex motion \cite{kwon1}.
Therefore, when there are multiple sources of hysteresis, a plot of $Q$ vs. $f^{-2}$ allows us to distinguish those sources.

We reveal the physical processes behind magnetic hysteresis through two types of experiments:
one is the standard magnetic hysteresis curves of $f$ and $Q$ as a function of a magnetic field,
and the other is a measurement of changes in $f$ and $Q$ after applying a high microwave current.
We call the latter \textit{current annealing}.
The crucial difference separating current annealing from an $M$-$H$ curve or its variants is that the height of the edge barrier (more precisely, the Bean--Livingston barrier \cite{bean1964}) is tuned by the microwave current, not by an external field \cite{tafuri2006}.
If a superconducting resonator is in a metastable state due to the edge barrier, the high microwave current can change the resonator's state via suppressing the edge barrier, resulting in different $f$ and $Q$ values.
Hence, this current annealing procedure allows us to investigate how the Meissner current and vortex configuration evolve as we tune the edge barrier without changing an external magnetic field.

\section{Methods}
\label{sec:method}

\begin{figure}
\centering
\includegraphics[scale=0.55]{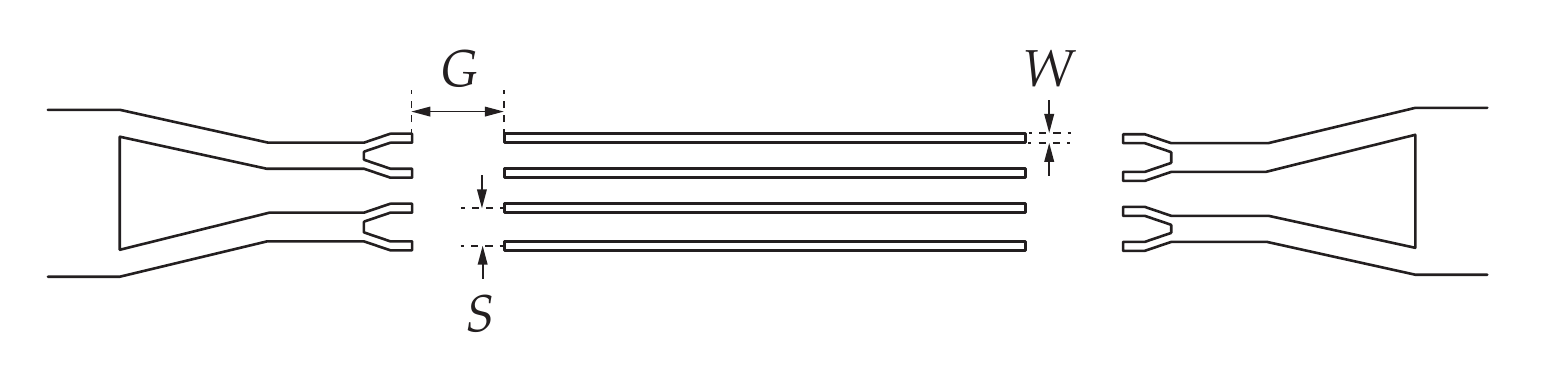}
\caption{\label{fig:config}
Geometry of the resonator.
$G$ is the gap between the feedline and the resonator. $W$ is the width of a strip. $S$ is the spacing between center of strips.
The value of $G$ is 350 $\mu$m; $W$, 15 $\mu$m; and $S$, 75 $\mu$m.
}
\end{figure}

The resonator used in this work is a trilayer superconducting microstrip resonator \cite{mohebbi2014}, whose film composition is Al/Nb/Al with the thickness is of 5/50/5 in units of nanometers.
The ground plane is made of a pure Nb layer of 50 nm in thickness.
All layers were magnetron sputtered under 5 mTorr Ar pressure.
The geometry of the resonator is shown in Fig.~\ref{fig:config}.
This is the same resonator discussed in Refs.~\cite{tang2017, kwon2}.
The details of the film growth conditions are reported in Ref.~\cite{tang2017}.

The measurements were performed in a dilution refrigerator.
A magnetic field perpendicular to the film $H_\perp$ was applied by tilting the resonator in a background magnetic field parallel to the microwave current $H_\textrm{bg}$ using a goniometer.
$H_\perp$ is obtained by $H_\perp = H_\textrm{bg} \sin\theta$.
A magnetic field parallel to the microwave current without a perpendicular component is denoted by $H_\parallel$.
In this work, $\mu_0 H_\textrm{bg} = 0.35$ T.
The details of the measurement configurations are described in Ref.~\cite{kwon1}.

Two different types of cooling procedure were used:
zero-field cooling (ZFC) and heat pulsing (HP).
For the ZFC procedure, the resonator is cooled without any magnetic field.
For the HP procedure, a heat pulse is applied to completely suppress superconductivity, then the resonator is cooled back in field to the target temperature.
The HP procedure is used to ensure a uniform vortex distribution and suppress the Meissner current as much as possible such that vortex motion becomes the dominant loss mechanism \cite{kwon1, song2009a}.
In this work, the default cooling procedure is ZFC; the HP data are labeled explicitly.
All measurements were made at 0.2 K.

\section{Hysteresis Curves}
\label{sec:hys}

\subsection{Expected Hysteresis Curves}
\label{sec:hys_theory}

Magnetic hysteresis effects in superconductors are determined by geometry, vortex pinning, and an edge barrier.
Regarding the geometry, our resonator is in the thin and wide strip limit:
$d \ll \Lambda \ll W$, where $d$ is the film thickness, $\Lambda$ is the screening length, and $W$ is the strip width.
Here, $\Lambda$ is given by $2\lambda \coth(d/\lambda)$ \cite{klein1990, irz1995, wei1996}, where $\lambda$ is the penetration depth.
For our resonator, $\lambda$ is about 200 nm \cite{kwon2}, which means $\Lambda \approx 1.4$ $\mu$m.
In this limit, the Bean--Livingston barrier is the dominant edge barrier; the geometrical barrier is unimportant.

As for vortex pinning and an edge barrier, we can consider two regimes based on the critical current density associated with vortex pinning $j_\textrm{p}$ and the depairing current density $j_\textrm{d}$.
In the strong pinning limit ($j_\textrm{p} \sim j_\textrm{d}$, Bean-type critical state models), the hysteresis is primarily due to vortex pinning \cite{brandt1993b, zeldov1994a}.
Because of the pinning potentials, vortices accumulate near the edges,
and the edge barrier is strongly suppressed \cite{kuznetsov1999}.
In the weak pinning limit ($j_\textrm{p} \sim 0$), vortex trapping due to an edge barrier is the main source of magnetic hysteresis \cite{schuster1994a, zeldov1994b, zeldov1994c, maksimov1995, maksimov1997}.
In this regime, vortices accumulate at the center because of the repulsive interaction between the Meissner current and vortices.
If $j_\textrm{p}$ is finite but significantly smaller than $j_\textrm{d}$, the number density of vortices is suppressed near both the edges and the center \cite{zeldov1994b, zeldov1994c, maksimov1995, elistratov1997, maksimov1998, elistratov2000, maksimova2001}.
Moreover, the edge barrier still contributes to the hysteresis significantly.
The generalized critical-state model describes the physical process in this regime \cite{maksimov1995, elistratov1997, maksimov1998, elistratov2000, maksimova2001}.

The resonator properties are determined by the complex resistivity $\rho_1 + \textrm{i}\rho_2$ and the microwave current density \cite{kwon1}.
Since most of the microwave current flows through the edges, the contribution from the screening current in those regions is amplified several orders of magnitude, and consequently, becomes comparable to the contribution from the rest of the resonator.\footnote{In this work, \textit{edge} means the region $W - \Lambda < |y| < W$, where $y$ is the distance from the center of the strip.
The \textit{screening current} mostly comprises the Meissner current, but also includes the supercurrent circulating around vortices.}
The screening current density near the edges $j_\Lambda$ contributes to the complex resistivity via quasiparticle generation;
the trapped flux $\Phi_\textrm{tr}$, via vortex motion \cite{kwon1}.
Thus, we start with the hysteresis curves of $j_\Lambda$ and $\Phi_\textrm{tr}$.
Based on those curves, we will construct expected hysteresis curves of $f^{-2}$ and $Q^{-1}$.
The reason for using those quantities is that the magnetic field dependences of $Q^{-1}$ and $f^{-2}$ are chiefly determined by $\rho_1$ and $\rho_2$, respectively \cite{kwon1}.

\begin{figure*}
\centering
\includegraphics[scale=0.50]{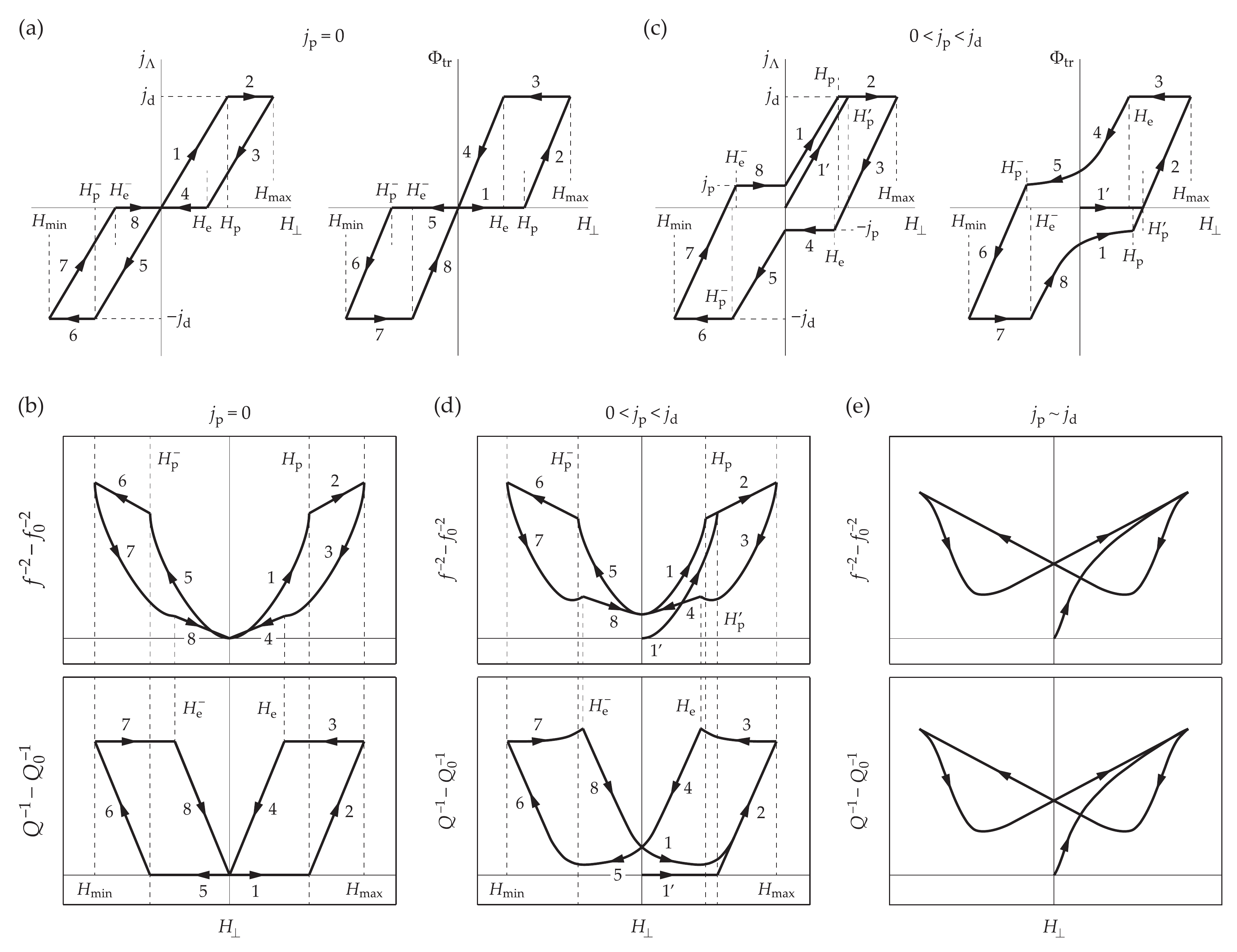}
\caption{\label{fig:hys_dia}
Schematic diagrams for magnetic hysteresis curves of various quantities.
(a) The screening current density near the edges $j_\Lambda$ and the trapped flux $\Phi_\textrm{tr}$ as a function of a magnetic field perpendicular to a film $H_\perp$ when $j_\textrm{p} = 0$.
(b) $f^{-2}$ and $Q^{-1}$ based on (a).
(c) $j_\Lambda$ and $\Phi_\textrm{tr}$ when $j_\textrm{p}$ is finite but still significantly lower than $j_\textrm{d}$.
(d) $f^{-2}$ and $Q^{-1}$ based on (c).
(e) $f^{-2}$ and $Q^{-1}$ in the strong pinning limit.
(a) and (c) are based on the generalized critical-state model \cite{maksimov1997, elistratov1997, maksimov1998, elistratov2000, maksimova2001}, whereas (e) is based on Bean-type models \cite{bothner2012b, brandt1993b, zeldov1994a}.
The numbers indicate the process order of hysteresis curves.
$H_\textrm{p}^{(-)}$ and $H_\textrm{e}^{(-)}$ are the (anti)vortex penetration and exit fields, respectively.
The prime in a letter or a number indicates the field or the process associated with the initial curve.
$j_\textrm{p}$ is assumed independent from $H_\perp$.
}
\end{figure*}


For simplicity, we will consider the weak pinning limit ($j_\textrm{p} = 0$) first.
Figure~\ref{fig:hys_dia}(a) shows schematic diagrams of $j_\Lambda$ and $\Phi_\textrm{tr}$ as a function of magnetic field perpendicular to a film $H_\perp$.
At first, only $j_\Lambda$ increases with $H_\perp$, and there is no vortex injection because of an edge barrier (process 1).
Once $H_\perp$ reaches the vortex penetration field $H_\textrm{p}$ at which $j_\Lambda$ becomes comparable to $j_\textrm{d}$, vortices begin to enter the film (process 2). In this process, $j_\Lambda$ remains largely unchanged.
As $H_\perp$ decreases, $j_\Lambda$ decreases, while the vortices are still trapped by the edge barrier (process 3).
The vortices start to be expelled from the film at the vortex exit field $H_\textrm{e}$ where $j_\Lambda$ is reduced sufficiently (process 4).
Processes 5--8 are identical to processes 1--4 except for the polarity of the vortices and the Meissner current.

Based on Fig.~\ref{fig:hys_dia}(a), we can construct the expected hysteresis curves of the quantities $f^{-2}$ and $Q^{-1}$ as shown in Fig.~\ref{fig:hys_dia}(b).
For process 1 ($0 \leq H_\perp < H_\textrm{p}$), $f^{-2}$ is completely determined by $j_\Lambda$ because of the absence of vortex penetration.
In this case, $f^{-2}$ varies quadratically with $H_\perp$ \cite{kwon1}, while $Q^{-1}$ changes little because the loss induced by quasiparticle generation is much lower than that induced by vortex motion in our experimental configuration \cite{kwon2}.
For process 2 ($H_\textrm{p} \leq H_\perp < H_\textrm{max}$), $Q^{-1}$ increases linearly with $H_\perp$ because vortices are accumulated near the center where the microwave current density distribution is roughly homogeneous;
$f^{-2}$ also increases linearly because of the inductive contribution of vortex motion.
For process 3 ($H_\textrm{e} < H_\perp \leq H_\textrm{max}$), $Q^{-1}$ does not change because $\Phi_\textrm{tr}$ is constant, whereas $f^{-2}$ drops quadratically.
For process 4 ($0 < H_\perp \leq H_\textrm{e}$), both $f^{-2}$ and $Q^{-1}$ decreases linearly.


Now, $j_\textrm{p}$ is finite but still significantly lower than $j_\textrm{d}$, such that a reasonably high edge barrier exists [Fig.~\ref{fig:hys_dia}(c)].
The generalized critical-state model predicts a couple of notable modifications \cite{elistratov1997, maksimov1998, elistratov2000, maksimova2001}:
(i) Trapped vortices start escaping when $j_\Lambda = - j_\textrm{p}$
(antivortices, when $j_\Lambda = j_\textrm{p}$).
(ii) $\Phi_\textrm{tr}$ is finite at zero field and remains finite until the next (anti)vortex penetration field.
(iii) During processes 3 ($H_\textrm{e} < H_\perp \leq H_\textrm{max}$) and 7 ($H_\textrm{min} \leq H_\perp < H_\textrm{e}^-$), some of the fluxes move towards the edges, although the total $\Phi_\textrm{tr}$ is still conserved.
This redistribution of vortices is often called \textit{flux defreezing} \cite{elistratov1997, maksimov1998, elistratov2000, maksimova2001}.

Applying those differences, the expected hysteresis curves of $f^{-2}$ and $Q^{-1}$ are obtained as shown in Fig.~\ref{fig:hys_dia}(d).
The differences from Fig.~\ref{fig:hys_dia}(b) are the following:
(i) $f^{-2} - f_0^{-2}$ always remains finite after the initial curve because of the inductive contribution from the motion of trapped vortices and the supercurrent circulating around those vortices.
(ii) Near the end of processes 3 and 7, $f^{-2}$ shows a small dip because $j_\Lambda$ passes zero just before $H_\textrm{e}^{(-)}$ [Fig.~\ref{fig:hys_dia}(c)]. 
(iii) $Q^{-1} - Q_0^{-1}$ is also always finite after the initial curve.
Although $\Phi_\textrm{tr}$ passes zero in processes 2 ($H_\textrm{p} \leq H_\perp < H_\textrm{max}$) and 6 ($H_\textrm{min} < H_\perp \leq H_\textrm{p}^-$), the number of vortices and antivortices is not zero because zero $\Phi_\textrm{tr}$ is due to a balance between vortices and antivortices.
(iv) $Q^{-1}$ increases during processes 3 and 7 until $H_\textrm{e}^{(-)}$.
This is due to the flux defreezing;
as vortices move towards the edges, they generate more loss because of the higher microwave current density near the edges.


Note that the rotation directions of $f^{-2}$ and $Q^{-1}$ at $H_\textrm{max}$ and $H_\textrm{min}$ are opposite in Fig.~\ref{fig:hys_dia}(b,d).
This behavior cannot be observed in Bean-type models that predict the same rotation direction for both quantities [Fig.~\ref{fig:hys_dia}(e)].
The reason is that vortices, rather than $j_\Lambda$, are responsible for changes in both $f^{-2}$ and $Q^{-1}$ \cite{bothner2012b}, and both $\rho_1$ and $\rho_2$ associated with vortex motion have essentially the same dependence on $H_\perp$ at sufficiently low temperatures, where the vortex creep contribution is negligible \cite{kwon1, pompeo2008}.

\subsection{Experimental Data}
\label{sec:hys_data}

\begin{figure*}
\centering
\includegraphics[scale=0.5]{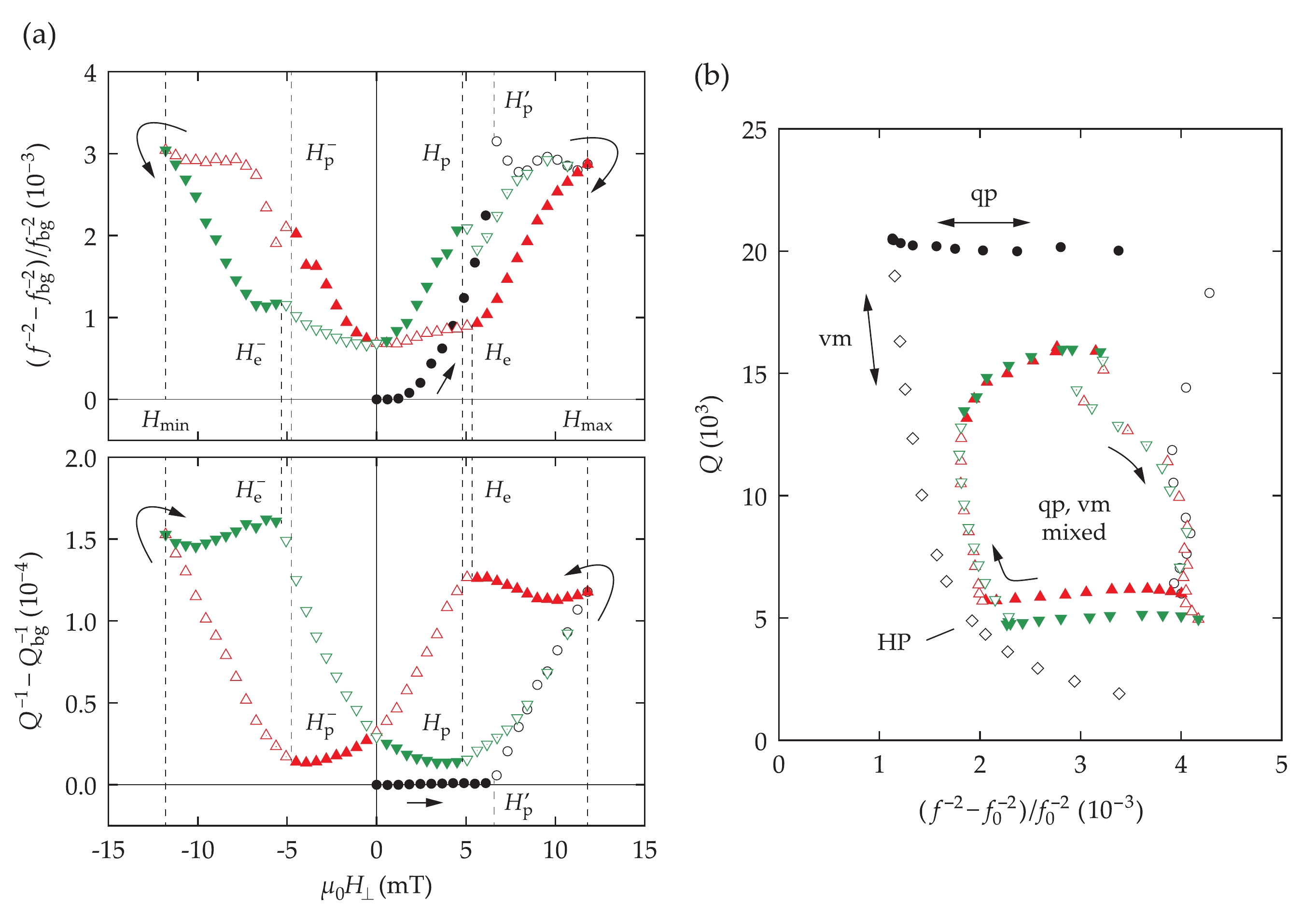}
\caption{\label{fig:hys}
(a) Hysteresis curves of $f^{-2}$ and $Q^{-1}$.
The subscript ``bg'' means that the quantity is measured with $H_\textrm{bg}$, but without tilting: $f_\textrm{bg} \equiv f(H_\parallel=H_\textrm{bg}, \theta=0)$ and $Q_\textrm{bg} \equiv Q(H_\textrm{bg}, 0)$.
$H_\textrm{p}^{(-)}$ is inferred based on the frequency anomalies;
$H_\textrm{e}^{(-)}$ is the field where $Q^{-1}$ starts to drop.
(b) A $Q$ vs. $f^{-2}$ plot of the data in (a).
The dominant loss mechanisms were determined by this plot and denoted by different symbols in both (a) and (b):
Filled symbols are the data whose loss is dominated by quasiparticle generation (qp);
empty symbols, dominated by vortex motion (vm);
and symbols with a dot, data where the two loss mechanisms are comparable.
The excitation power from the vector network analyzer was $-40$ dBm.
Errors are comparable to or smaller than the size of symbols.
The HP data is from Ref.~\cite{kwon2}.
}
\end{figure*}

Figure~\ref{fig:hys}(a) shows the magnetic hysteresis curves of $f^{-2}$ and $Q^{-1}$.
Note that it shows all important features predicted in Sec.~\ref{sec:hys_theory}:
opposite rotation of $f^{-2}$ and $Q^{-1}$ at $H_\textrm{max}$ and $H_\textrm{min}$ (indicated by black curved arrows),
small dips on $f^{-2}$ just before $H_\textrm{e}^{(-)}$, and
an enhancement of $Q^{-1}$ before $H_\textrm{e}^{(-)}$.
These confirm the validity of the generalized critical-state model for our resonator.

There are also a couple of differences.
First, there are anomalies (a peak followed by a dip) in the $f^{-2}$.
In our previous work, we showed that these frequency anomalies are an indication of release of the Meissner current due to the repulsive interaction between the Meissner current and vortices \cite{kwon1}.
Hence, we define the fields at which those frequency anomalies appear as $H_\textrm{p}^{(-)}$ and $H_\textrm{p}'$.

Secondly, the frequency anomaly associated with $H_\textrm{p}^{(-)}$ appeared earlier than was expected in Fig.~\ref{fig:hys_dia}(d), and was followed by a considerable increase in $f^{-2}$ (symbols with a dot).
This indicates that (anti)vortices start penetrating before $j_\Lambda$ meets $\pm j_\textrm{d}$.
This is likely due to trapped (anti)vortices.
The vortex penetration field associated with the Bean--Livingston barrier is determined by a balance between two interactions;
the repulsive interaction between the Meissner current and vortices, and the attractive interaction between vortices and (image) antivortices outside of a superconductor \cite{matsushita}.
If there are a number of antivortices trapped inside of the superconductor, an additional attractive interaction between vortices and the trapped antivortices reduces the vortex penetration field.
Once vortices penetrate, some of the vortices and antivortices meet and annihilate each other;
$j_\Lambda$ can then increase further.

Lastly, after changing the field sweep direction at $H_\textrm{max}$ and $H_\textrm{min}$, $Q^{-1}$ decreases slightly.
During that process, the slope of $f^{-2}$ is less than that expected from quadratic dependence on $H_\perp$.
This indicates that some vortices are pinned near the edges.
Since those vortices are close to the edges and not trapped by the edge barrier, they disappear quickly once we change the sweep direction.


From the discussion so far, we see that the heart of magnetic hysteresis curves in a superconducting device with a high surface barrier is the interchanging role between $j_\Lambda$ and $\Phi_\textrm{tr}$: when $j_\Lambda$ changes, $\Phi_\textrm{tr}$ stays similar and vice versa [Fig.~\ref{fig:hys_dia}(a,c)].
This suggests that revealing the dominant physical process for each part of the hysteresis curve is crucial for understanding magnetic hysteresis effects.
For this, we use a plot of $Q$ vs. $f^{-2}$.

Note that, in this plot [Fig.~\ref{fig:hys}(b)], the hysteresis curves in Fig.~\ref{fig:hys}(a) form loops rotating clockwise.
If there was only one source of hysteresis, then the $Q$ vs. $f^{-2}$ curves would collapse to a single line regardless of the existence and details of hysteresis because the characteristic relation between the real and imaginary parts of the complex resistivity would be preserved throughout the measurement.
Hence, the existence of loops, rather than just a line, shows that multiple sources are contributing to the hysteresis.
In this case, they are the Meissner current along the edges of the microstrips and vortex pinning.

Also note that, a process of varying quasiparticle numbers evolves horizontally keeping $Q^{-1}$ constant,
while a process of varying vortex numbers evolves as a nearly vertical line in a plot of $Q$ vs. $f^{-2}$ [denoted by arrows in Fig.~\ref{fig:hys}(b)].
The reason is that, for our device in the range of magnetic fields studied, quasiparticle generation is a very inductive process such that $Q$ is dominated by vortex motion \cite{kwon2}.
The dominant physical process of each part of the hysteresis curve (denoted by various symbols) were identified in this plot and found to be consistent with those described in Fig.~\ref{fig:hys_dia}(c).

\section{Current Annealing}
\label{sec:ca}


\begin{figure}
\centering
\includegraphics[scale=0.5]{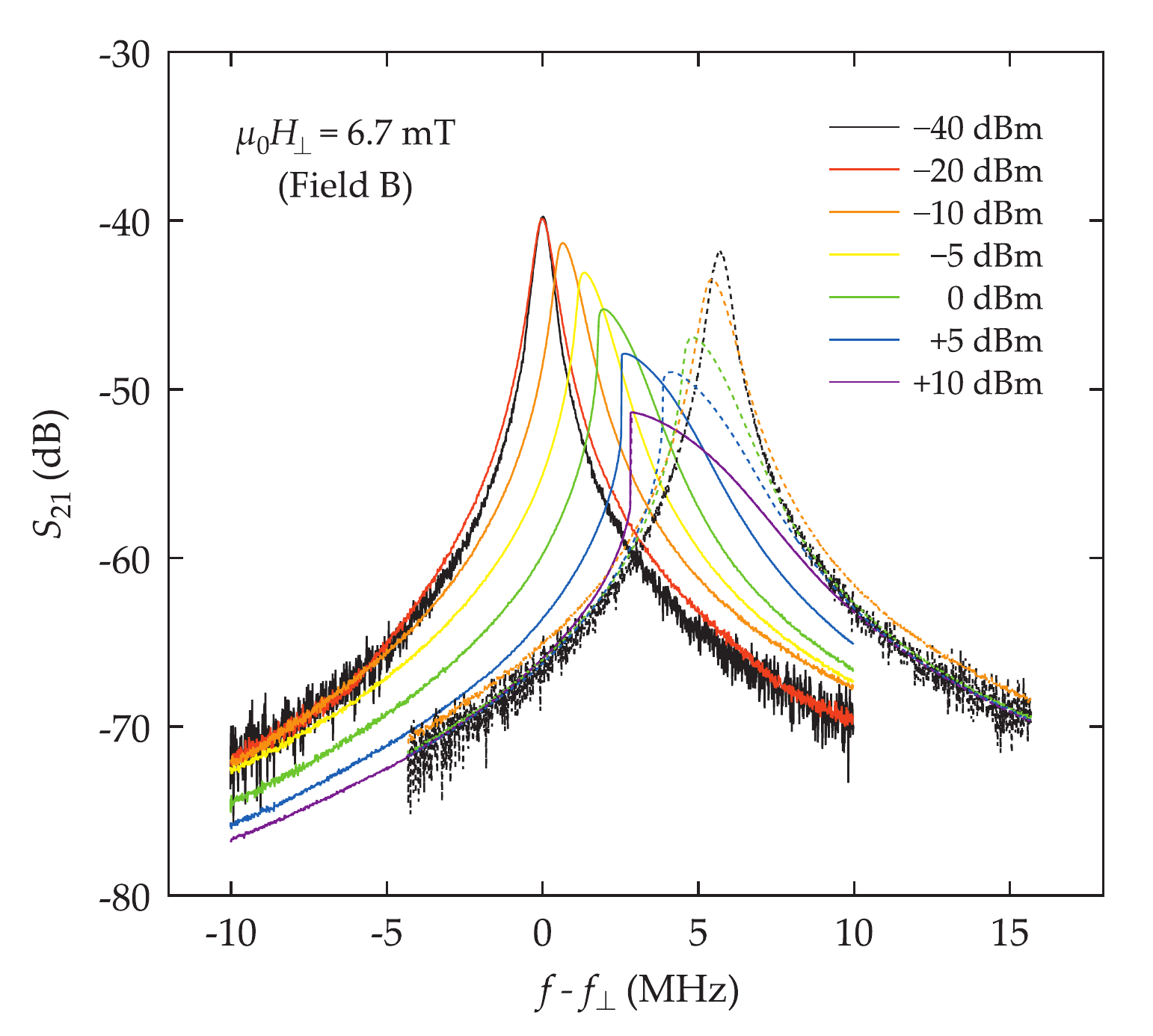}
\caption{\label{fig:ca_spec}
$S_{21}$ curve shift during the current annealing.
Annotated powers ($P_\textrm{VNA}$) were applied sequentially.
$f_\perp$ is the resonance frequency before the current annealing.
Solid lines are the results from the initial current annealing and dashed lines are from the second annealing. 
Results from further sequences are identical to the second one.
The sweep direction is from low to high frequency.
}
\end{figure}

The current annealing measurements were done by the following procedure:
First, $H_\perp$ is applied after ZFC.
Then the excitation power from the vector network analyzer $P_\textrm{VNA}$ is applied from low power ($-40$ dBm) to the power called the current annealing power $P_\textrm{CA}$.
Note that as $P_\textrm{VNA}$ increases, the spectrum not only shows the Duffing-like nonlinearity but also shifts to a higher frequency as shown in Fig.~\ref{fig:ca_spec} (solid lines).
Once $P_\textrm{VNA}$ reaches $P_\textrm{CA}$ ($+10$ dBm for Fig.~\ref{fig:ca_spec}),
$P_\textrm{VNA}$ is reduced to $-40$ dBm.
We call the procedure up to this point the initial current annealing.
After the initial current annealing, no spectrum shift is observed for $P_\textrm{VNA} \leq P_\textrm{CA}$ (dashed lines).
The position of the spectrum from a low-power measurement is completely determined by the highest power prior to the low-power measurement regardless of a history of $P_\textrm{VNA}$.


\begin{figure*}
\centering
\includegraphics[scale=0.5]{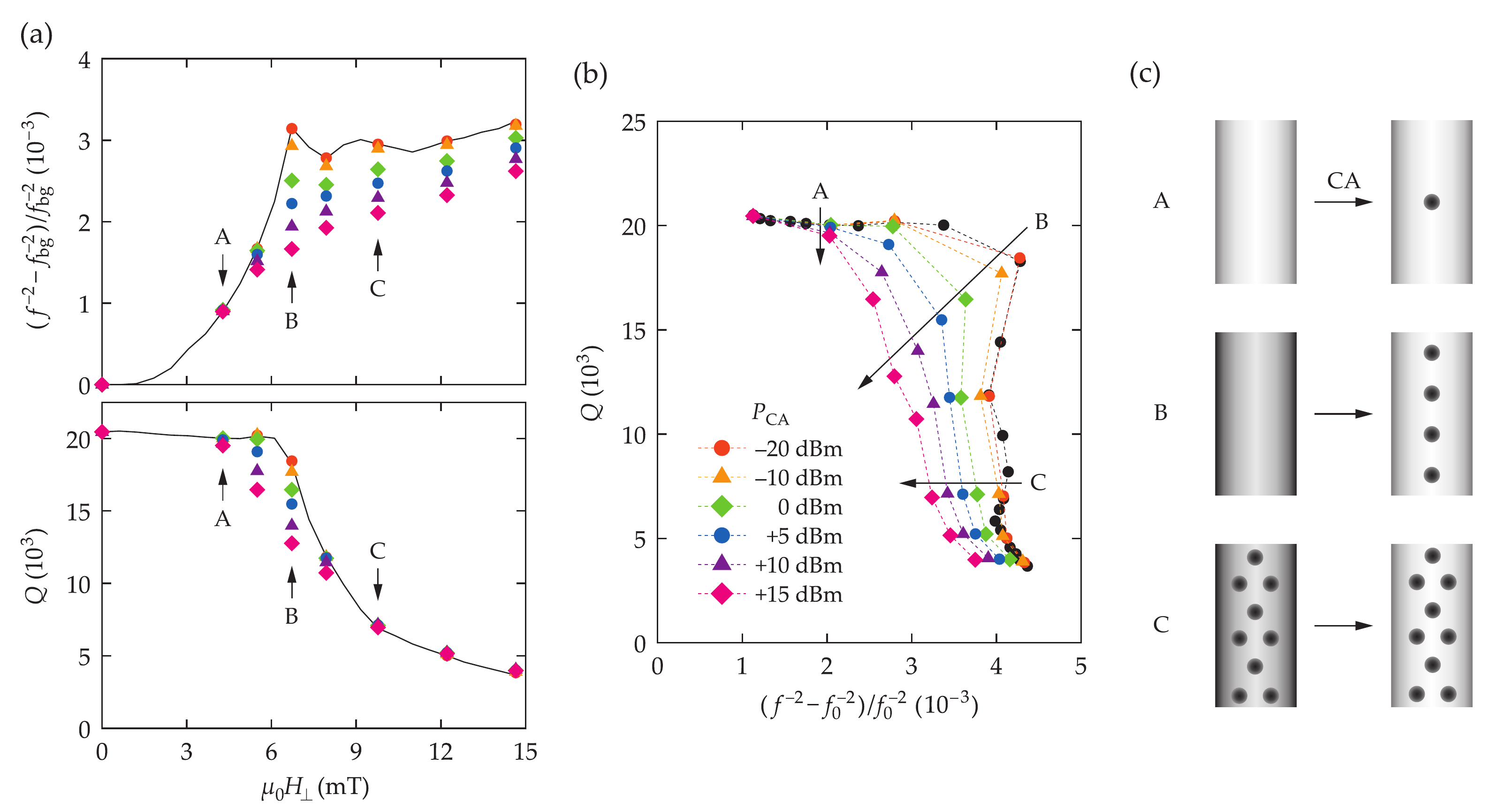}
\caption{\label{fig:ca}
(a) Final $f^{-2}$ and $Q$ measured with $P_\textrm{VNA} = -40$ dBm after the initial current annealing.
(b) A $Q$ vs. $f^{-2}$ plot of the data in Fig.~\ref{fig:ca_spec}.
In (a,b), symbols with different colors mean that the data were taken with a different $P_\textrm{CA}$,
while all black lines/symbols were taken without current annealing and are from Ref.~\onlinecite{kwon2}.
Arrows in (b) indicate the direction of evolution by the current annealing.
Dashed lines in (b) are guides to the eye.
(c) The Meissner current and vortices configuration before and after the current annealing at the designated fields.
The gray gradient indicates the schematic distribution of the Meissner current density; the darker area is higher current density. The dark gray circles are vortices.
}
\end{figure*}

Each $H_\perp$ yields different values of $f^{-2}$ and $Q$ after the initial current annealing, as shown in Fig.~\ref{fig:ca}(a).
Note that the frequency anomaly becomes weaker as $P_\textrm{VNA}$ increases, indicating the suppression of the edge barrier.
Again, plotting $Q$ vs. $f^{-2}$ [Fig.~\ref{fig:ca}(b)] reveals the physical processes behind Fig.~\ref{fig:ca}(a).
Like the arrows in Fig.~\ref{fig:hys}(b), arrows in Fig.~\ref{fig:ca}(b) suggest the following processes occur during current annealing [Fig.~\ref{fig:ca}(c)]:
(i) At the field A, the edge barrier is suppressed by the microwave current; consequently, vortices penetrate into the resonator.
The number of newly injected vortices is not, however, enough to yield a notable change in the Meissner current.
(ii) At the field B, a large number of vortices penetrate and reduce $Q$;
the Meissner current is expelled by the interaction with the newly injected vortices, resulting in the reduction of $f^{-2}$.
(iii) At and above the field C, the number of vortices stays similar because the edge barrier is already completely suppressed by $H_\perp$;
the Meissner current is expelled notably.
This is due to the enhancement of the kinetic energy of existing vortices by the microwave current, or a ``shaking'' of the vortices.
The displacement of vortices during the shaking expels the Meissner current.

\section{Conclusion}

In conclusion, we explored magnetic hysteresis in a superconducting microstrip resonator with a high edge barrier.
We analyzed two types of magnetic hysteresis effects:
one appears while sweeping a perpendicular magnetic field, and the other appears while tuning the edge barrier by high microwave current (current annealing).
We found that the hysteresis in a device with a high surface barrier is qualitatively different from that in the strong pinning limit and can be understood based on the generalized critical-state model.
In particular, we revealed the physical processes behind magnetic hysteresis using a plot of $Q$ vs. $f^{-2}$.
By doing this, we found that the interplay between the Meissner current and vortex pinning is crucial for understanding hysteretic behaviors.
We believe our method is applicable to other planar superconducting resonators.

\begin{acknowledgements}
This work is supported by the Canada First Research Excellence Fund, 
the Canada Excellence Research Chairs (grant No. 215284), 
the Natural Sciences and Engineering Research Council of Canada (grant Nos. RGPIN-418579 and RGPIN-04178), 
and the Province of Ontario.
The University of Waterloo's Quantum NanoFab was used for this work. 
This infrastructure is supported by
the Canada Foundation for Innovation, 
the Ontario Ministry of Research \& Innovation, 
Industry Canada, and 
Mike \& Ophelia Lazaridis. 
\end{acknowledgements}

\end{document}